\documentclass[reprint, amsmath,amssymb,aps,pra,longbibliography]{revtex4-1}
\usepackage[utf8]{inputenc}
\usepackage{graphicx}
\usepackage{dcolumn}
\usepackage{bm}%
\usepackage[colorlinks=true ,citecolor=blue]{hyperref}
\usepackage[english]{babel}
\usepackage{endnotes}
\usepackage{accents}
\usepackage{makeidx}
\usepackage{setspace}
\usepackage{braket}
\usepackage[makeroom]{cancel}
\usepackage[utf8]{inputenc}

\def\brack#1{ \left(   #1 \right) }
\def\curlbrack#1{\left\{ #1 \right\}}
\def\blockbrack#1{ \left[   #1 \right] }

\usepackage{color}

\begin{document}
\title{Quantizing L\'evy flights}
\author{R. A. Mulder,$^{1,2}$ M. A. Caracanhas,$^{1,3}$ and C. Morais Smith$^1$}
\affiliation{$^1$Institute for Theoretical Physics, Utrecht University, 
The Netherlands. \\
$^2$ Department for HPS, Cambridge University, Trinity College, United Kingdom.\\
$^3$ Instituto de F\'\i sica de S\~{a}o Carlos, Universidade de S\~{a}o Paulo, Brazil. }

\begin{abstract}
The Caldeira-Leggett model of quantum Brownian motion is generalized using a generic velocity-dependent coupling. That leads to the description of a set of models able to capture Markovian and non-Markovian versions of Brownian and L\'evy statistics, depending on the functional form of the coupling and on the spectral function of the reservoir. One specific coupling force is found that establishes a connection with L\'evy statistics of cold atoms in \textit{Sisyphus} laser cooling. In the low-velocity limit, this also gives rise to additional inertia of the Brownian particle, reproducing the Abraham-Lorentz equation from first principles for a superohmic bath. Through path-integral quantization in Euclidean time, the environment is integrated out, leaving a set of non-local effective actions. These results further serve as starting points for several numerical calculations, particularly decoherence properties of non-ohmic baths.
\end{abstract}

\date{\today}
\maketitle

\section{Introduction}
The description of  open quantum systems is often a remarkable challenge. A paradigmatic example is Brownian motion, the random-walk behavior exhibited by a particle when subjected to a bath composed of smaller particles. Its classical trajectory is described by the Langevin equation, which has a velocity-dependent term that makes direct canonical quantization problematic since energy conservation is violated. To solve this problem, Caldeira and Leggett \cite{CaldLegg,Caldeira} modeled the bath as a collection of harmonic oscillators, which allowed them to close the system and investigate quantum aspects of Brownian motion to describe the quantum flux dynamics in Josephson junctions and SQUIDs \cite{Caldeira,CaldLegg2,CaldLegg3}.

\begin{figure}[tbh]
\centering
\includegraphics[width=0.5\textwidth]{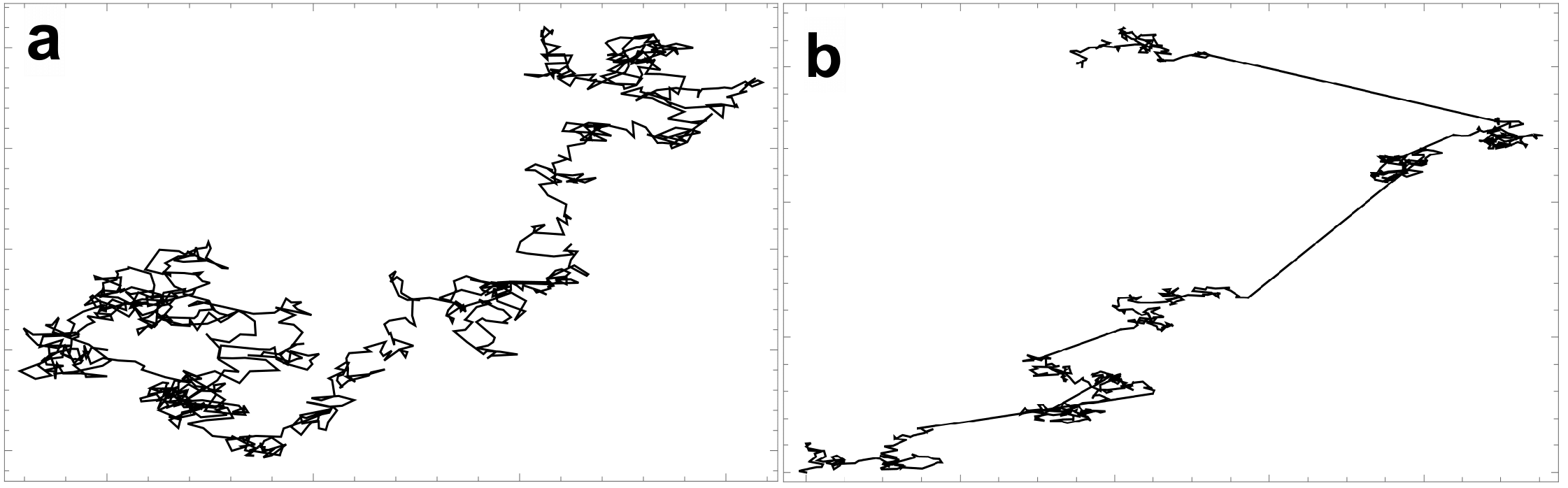}
\caption{Comparison between (a) Brownian motion and (b) L\'evy motion. For L\'evy motion, long steps are more frequent and make the dominant contribution to transport.}
\label{fig:1}
\end{figure}

Many simple physical systems adhere to the random-walk behavior of Brownian motion. Nevertheless, more complex motions are also found in nature. A typical example is L\'evy motion, which corresponds to a random walk for most of the time, but has occasional large `jumps', the so-called `L\'evy walks' \cite{Zaburdaev} (see Fig.~\ref{fig:1}). The L\'evy distribution is a generalization of the Gaussian distribution,
characterized by power-law tails and divergent moments. L\'evy statistics is for example applied to model turbulent flow \cite{West}, in hopping processes in polymer physics \cite{Rice}, or in laser cooling experiments \cite{Barthelemy,Bardou1}.
 L\'evy statistics is also connected to diffusion in a  Sierpinski fractal  \cite{Kravtsov}. Recently, a quantum fractal has been experimentally realized in the nano-domain and the electronic wavefunctions were shown to experience the fractal dimension of the Sierpinski gasket, d=1.58 \cite{Kempkes}. An important question in this context is then how to model a {\it quantum L\'evy system}.

Here we generalize the Caldeira-Leggett model to answer this question. First, we change the interaction between the Brownian particle and the bath to depend on the velocity of the particle in a generic way. A second tool to describe different types of processes is the specific form of the spectral function ${J} (\omega)$, which characterizes the reservoir. In general, ${ J} (\omega) = \eta {\omega}^s$, with $s=1$ describing ohmic processes and $s < 1$ or $s > 1$ describing subohmic and superohmic baths, respectively. As an application of our generic model we have chosen a particular non-polynomial coupling which, together with an ohmic bath to describe a Markovian process, reproduces a force that gives rise to L\'evy walks as encountered in ultracold-atoms experiments, e.g. \textit{Sisyphus} laser cooling  \cite{Lutz,Afek}. In the low-velocity regime, in contrast, we find an equation that resembles the Abraham-Lorentz equation in the superohmic regime, i.e., the equation for the self-interaction of an electron with its own radiation field  \cite{Barone}.

Further, we use path-integral quantization to construct a quantum version for the generalized velocity-dependent model. After tracing out the bath, it turns out to be possible to find an effective action without a special choice for the form of the coupling; hence it remains completely general. Reducing to the linear-velocity case, we obtain an effective action that can be interpreted as a bath-induced resistance to a change in velocity of the Brownian system---on top of classical inertia terms. These quantum models directly correspond to models of the generalized Langevin equation, i.e. the Brownian and L\'evy regimes, depending on the choice for the coupling force. 


\section{The L\'evy distribution} \label{LaserCool}
\begin{figure}
\centering
\includegraphics[width=0.49\textwidth]{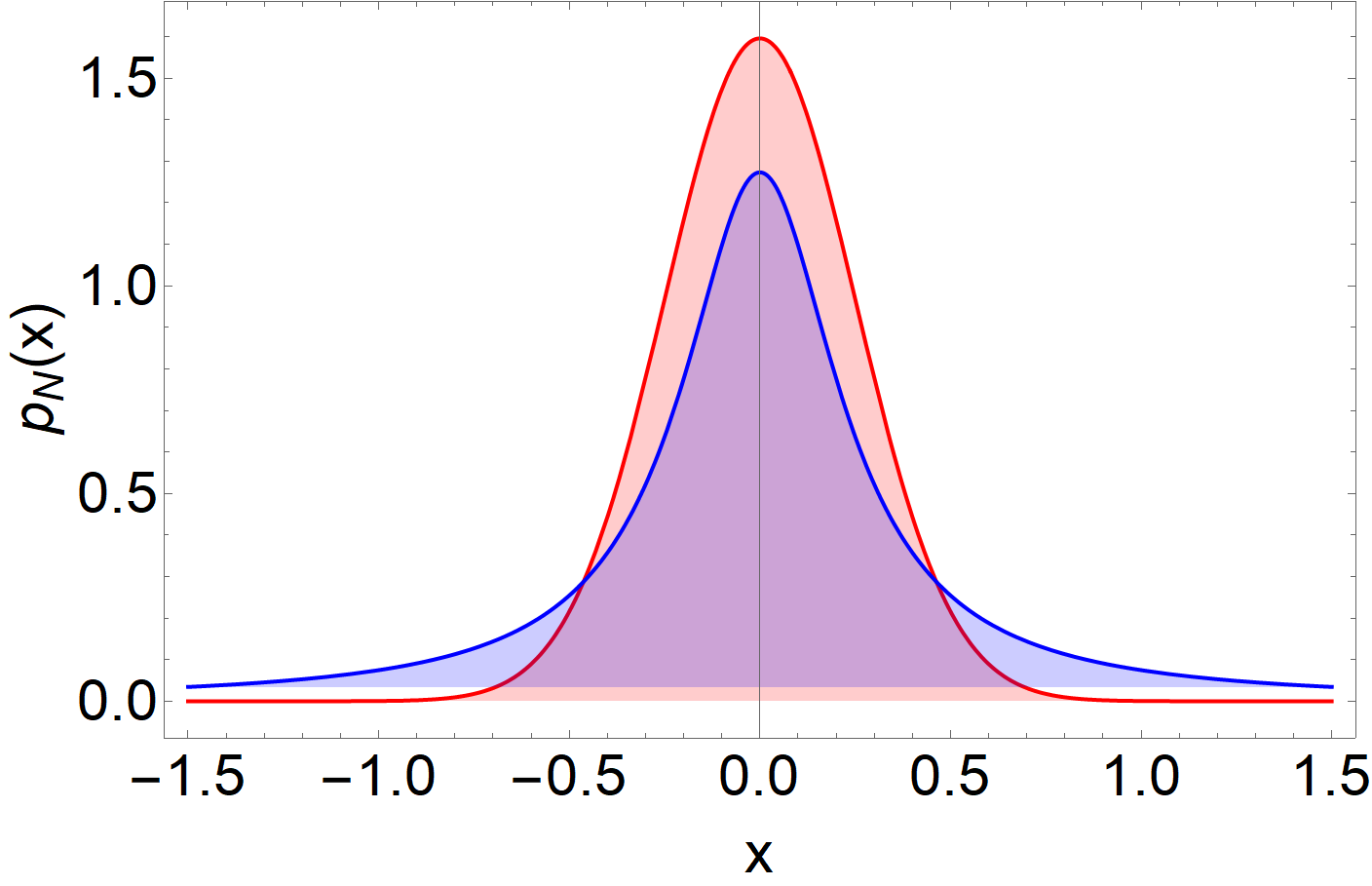}
\caption{Sketch of the Gaussian distribution (red) and of the L\'evy distribution (blue). Because the support of the L\'evy distribution is decaying at a lower rate than that of the Gaussian curve, a property called `heavy tails', the probability for a step deviating much from the mean is higher. Such a highly-deviating step, then, corresponds to an event occurring in the tails of L\'evy distribution.}
\label{fig:peak}
\end{figure}

\noindent In 1937, the French mathematician Paul L\'evy tried to find solutions to the question \cite{Levy}
\begin{quote}
When does the probability  $P_N(x)$ for the sum of $N$ steps $x=x_1+x_2+...+x_N$ has the same distribution (up to a normalization) as for the individual steps $p_i(x), \;i=1...N$ ?
\end{quote} The trivial solution is the Gaussian probability distribution
\begin{equation}
P(x)=\frac{1}{\sqrt{2\pi \sigma^2}} \exp\left(\frac{1}{2\sigma^2}x^2\right),
\end{equation}
for $\sigma$ the standard deviation, since the sum of $N$ Gaussian distributions is again a Gaussian. However, L\'evy found that there are additional solutions. These are called L\'evy distributions, which have the following form in Fourier space 
\begin{equation}
P_N(k)=\exp(-Na|k|^{\beta}),
\end{equation} where $N$ represents the total number of steps.
For $\beta=2$ and $a=\sigma^2/2$, we have a Gaussian distribution. We calculate its form in $x$-space with an inverse Fourier transform, which should return a Gaussian distribution,
\begin{align}
P_N(x)&=\int\limits_{-\infty}^{\infty} \frac{dk}{2\pi}
\exp\left(\frac{-N\sigma^2}{2}|k|^2+ikx\right) \\ \nonumber
&= \frac{1}{\sqrt{2\pi N \sigma^2}}\exp\left(-\frac{x^2}{2 N \sigma^2}\right),
\end{align}
as anticipated. This results in the random-walk behavior characteristic of Brownian motion. For $\beta=a=1$, however, we have the Cauchy-Lorentz or L\'evy distribution, which, transformed back to $x$-space reads
\begin{align}
P_N(x)=\int\limits_{-\infty}^{\infty}\frac{dk}{2\pi}\exp(-N|k|+ikx)  =\frac{1}{\pi}\frac{N}{N^2+x^2}.
\end{align}
A comparison between the Gaussian and the L\'evy distribution can be observed in Figure \ref{fig:peak}.
 The L\'evy distributions are the generalization of the Gaussian distribution in the sum of large numbers of independent variables in cases where the variances of the variables diverge, which is known as the generalized central limit theorem. In contrast to the Gaussian-distributed momentum of Brownian motion, the L\'evy distribution has `heavy tails', which increase the probability for making long jumps \cite{Klafter}.

L\'{e}vy statistics is now recognized as a good tool for studying many anomalous diffusion problems where standard statistics fail. There are plenty of experiments performed in the ultracold atoms community which allow us to study the anomalous diffusion with samples displaying non-Boltzmann-Gibbs statistics \cite{Katori,Bardou2,Sagi,Barkai}. The signature of anomalous diffusion is given by the presence of a power-law tail in the atomic momentum distribution, which can be verified through the absorption image of the free expanding atom cloud after being released from the confinement trap.  

Here, we will discuss the technique of Sisyphus laser cooling (sometimes called polarization gradient cooling), which is a type of laser cooling that allows atoms to reach temperatures below the Doppler cooling limit \cite{Peter}. This technique consists of a one-dimensional optical lattice formed by two counter-propagating laser beams with linear perpendicular polarization. The cooling and confinement rely on velocity-dependent or position-dependent absorption of laser photons followed by spontaneous emission. The complexity in the absorption-emission process can, theoretically, be quantitatively understood in terms of non-ergodic random processes dominated by a few rare events, such as the recoil after spontaneous emission; Brownian as well as anomalous diffusion such as L\'evy walks can be observed, depending on the heat bath implemented by the field of the lattice lasers, which  determines both damping and fluctuations in the transport equation \cite{Marksteiner}. In the context of the L\'evy walks in Sisyphus cooling, the following momentum-dependent force  was recently proposed,
\begin{equation} \label{Marksteiner}
f(p)=\frac{p}{1+p^2/p_c^2},
\end{equation} with a momentum cut-off $p_c$ \cite{Marksteiner, Kessler}. This is the force that we shall discuss and reproduce in Section \ref{Extending}.

\section{The Caldeira-Leggett model: Bilinear Coordinate-Coordinate coupling} \label{CLsection}

The framework for Brownian motion makes it difficult to find a theory of \textit{quantum} Brownian motion via conventional quantization techniques. The reason is that either the Lagrangian (for path-integral quantization) or the Hamiltonian (for canonical quantization) will have an explicit time-dependence in order to reproduce the velocity-dependent friction term of the Langevin equation. Hence, energy is not conserved and, although the non-conservation of energy is natural for open systems, this makes direct quantization impossible. In addition, this framework is a phenomenological one, i.e. the diffusion constant or the viscosity can only be determined experimentally for different materials that the environment can consist of. This is to be expected, since a microscopic description of the environment was absent in the first place. To go beyond phenomenology, one needs a theory where the origin of the viscosity of a particular medium can be explained on a theoretical basis. Yet there is no action principle that allows for the derivation of the phenomenological Langevin equation solely in terms of the Brownian particle. Thus, we will have to look for a microscopic description of the bath.

The gap between phenomenological modeling and precise microscopic formulation is bridged by the Caldeira-Leggett model \cite{CaldLegg}. It describes both the system and the bath, and their interaction by one Lagrangian, while the physical origin of this interaction is initially kept unspecified for generality. The bath is modeled as a collection of harmonic oscillators, linearly coupled to the system of interest. This is a \textit{minimal model}, since every perturbation of the bath can be approximated by a harmonic potential. Hence, the validity of this model is restricted to weak perturbations of the bath. Note that this does not imply that the induced dissipation  is necessarily weak: the large number of environmental degrees of freedom guarantees that we can describe strongly dissipative systems. The Lagrangian reads
\begin{equation}
\mathcal{L}=\mathcal{L}_S+\mathcal{L}_B+\mathcal{L}_I+\mathcal{L}_{C.T.},
\end{equation}
where the abbreviations stand for `system', `bath', `interaction' and `counterterm', respectively.
  
For simplicity, we let the system be described by one generalized coordinate $Q$. In one dimension, the Lagrangian of a particle of mass $M$, subject to an external potential $V(Q)$, reads
\begin{equation} \label{LagSys}
\mathcal{L}_S=\frac{1}{2}M\dot{Q}^2-V(Q).
\end{equation}
The bath is modeled as a collection of $N$ harmonic oscillators, labeled by an index $k$, with masses $m_k$, coordinates $q_k$, and at natural frequencies $\omega_k$,
\begin{equation}
\mathcal{L}_B=\frac{1}{2}\sum\limits_{k=1}^Nm_k\dot{q}^2_k -\frac{1}{2}\sum\limits_{k=1}^Nm_k \omega_k^2 q_k^2.
\end{equation}
In principle, the number $N$ of harmonic oscillators is very large, such that we can safely work within the thermodynamic limit, i.e. we can let $N \rightarrow \infty$ at the end of the calculation.

The interaction is of the coordinate-coordinate type, which means that the system coordinate $Q$ is linearly coupled to each oscillator coordinate $q_k$, as if attached to a spring, with `spring constants' $C_k$,
 \begin{equation}
 \mathcal{L}_I=Q\sum\limits_{k=1}^N C_kq_k.
 \end{equation}
 This interaction is the simplest one to write down (remember this is a minimal model), but it turns out to be quite general \cite{CaldLegg}. In principle, the $C_k$ should be seen as negative constants, since springs tend to \textit{restore} extensions.
Caldeira and Leggett also include a counterterm,
\begin{equation} \label{counterterm}
\mathcal{L}_{C.T.}=-\frac{1}{2}Q^2\sum\limits_{k=1}^N \frac{C_k^2}{m_k\omega^2_k},
\end{equation}
which depends on the parameters of the environment, but not on its dynamical variables. This renormalization term ensures that the minimum of the effective potential remains centered about the \textit{bare} potential $V(Q)$.

Having discussed the individual components of the Lagrangian, we can write the Caldeira-Leggett Lagrangian:

\begin{eqnarray} \label{CL} \nonumber 
\mathcal{L}&=&
\underbrace{
\frac{1}{2}M\dot{Q}^2-V(Q)}_{\text{System}}
+
\underbrace{
\frac{1}{2}\sum\limits_{k=1}^Nm_k(\dot{q}^2_k-\omega_k^2 q_k^2)
}_{\text{Bath}} \\ 
&+&
\underbrace{
Q\sum\limits_{k=1}^N C_kq_k
}_{\text{System/Bath-INT.}}
-
\underbrace{
\frac{1}{2}Q^2\sum\limits_{k=1}^N \frac{C_k^2}{m_k\omega^2_k}
}_{\text{Counterterm}}.
\end{eqnarray} From this Lagrangian, one can derive the equation of motion for the system, which reproduces the Langevin equations after the bath coordinates are eliminated. 

To successfully describe the bath, one needs to specify a continuous frequency density distribution, instead of the discrete oscillator distribution. This is called the `spectral function' $J(\omega)$, which  should arise from the additional information specified by the microscopics of the bath constituents \cite{Weiss},
\begin{equation} \label{SD}
J(\omega)=\frac{\pi}{2}\sum\limits_k\frac{C_k^2}{m_k\omega_k}\delta(\omega-\omega_k),
\end{equation}
where the factor $\pi/2$ is put in for later convenience. This form follows from the Kubo formula of linear response theory, and it models the linear response of the bath coordinates $q_i$ to a perturbation. The form of the spectral density in Eq.~\eqref{SD} allows one to convert discrete sums over oscillators into continuous integrals over frequency,
\begin{equation} \label{useJ}
\sum\limits_{k} \frac{C_k^2}{m_k \omega_k^2} \cos[\omega_k(t-t')]=\frac{2}{\pi}\int\limits_0^{\infty}d\omega \frac{J(\omega)}{\omega}\cos[\omega(t-t')].
\end{equation}

Now, one assumes a specific form of the spectral function, which falls apart into three classes, namely the ohmic, subohmic, and superohmic cases, which are linear, sublinear or higher polynomials in frequency, respectively \cite{CaldLegg}. Also, the spectral density vanishes for $\omega>\Omega$, i.e. a certain high-frequency cut-off $\Omega$, which fixes the timescale of the problem and is therefore inversely proportional to the relaxation time $\tau^{-1}$. Hence, one writes
\begin{equation} \label{ssohmic}
J(\omega) = \eta \omega^s \hspace{2cm}
\left\{
	\begin{array}{ll}
		\text{subohmic,}  & \mbox{if } s<1 \\
		\text{ohmic,}	  & \mbox{if } s=1 \\
		\text{superohmic,}	& \mbox{if } s>1
	\end{array}
\right.
\end{equation}
where $\eta$ is a proportionality constant, which plays a phenomenological role here. As indicated in the above, this phenomenological input is necessary to reproduce the (phenomenological) Langevin equation of the open system approach. The next step that we take is to generalize the model even further, namely to a generic velocity coupling.

\section{Generalized Caldeira-Leggett Model: velocity-coupling to the bath} \label{Extending}

 In this section, we will generalize the Caldeira-Leggett model from a coordinate-coordinate coupling to a coupling that is a general function of the velocity of the Brownian particle multiplied by the coordinates of the bath oscillators.  The  motivation to choose such a velocity-dependent coupling is to reproduce the force $p/(1+p^2)$ in the approach to L\'evy motion in Sisyphus cooling of Marksteiner, Ellinger and Zoller \cite{Marksteiner}---as explained in Section \eqref{LaserCool}. Realizing that the general-coordinate coupling would not give us the desired result, we chose to proceed with a completely general coupling to velocity. It turns out that such an approach allows one to reproduce the desired force in terms of velocity by choosing a specific coupling that will give us the intended result; and, hence, to construct a closed effective model exhibiting L\'evy behavior.

As a first step, we expand upon  the coordinate-coordinate coupling of the Caldeira-Leggett model \cite{CaldLegg} and replace it with a coupling of a general function of the velocity of the Brownian particle $F[\dot{Q}]$ to each of the coordinates of the bath, assuming a separable interaction $F_i[\dot{Q}]=\tilde{C}_iF[\dot{Q}]$. In this case,  the coupling constants $\tilde{C}_i$ are not interpreted as spring constants, but have a dimension depending on the particular realization of $F[\dot{Q}]$. The counterterm will also acquire a different physical interpretation. For linear-velocity coupling, e.g. it will renormalize the mass of the Brownian particle. 

The Lagrangian of the model then reads
\begin{widetext}
\begin{eqnarray} \label{CLV0}
\mathcal{L} =
\frac{1}{2}M\dot{Q}^2-V(Q)
+
\frac{1}{2}\sum\limits_{k=1}^Nm_k(\dot{q}^2_k-\omega_k^2 q_k^2) 
+
F[\dot{Q}]\sum\limits_{k=1}^N \tilde{C}_kq_k
-
\sum\limits_{k=1}^N\frac{\tilde{C}_k^2}{2m_k\omega_k^2}F^2[\dot{Q}],
\end{eqnarray}
\end{widetext} 
where $M$ is the mass of the particle, $Q$ its coordinate and the reservoir is characterized by harmonic oscillators with mass $m_k$, frequency $\omega_k$, coordinates $q_k$, and dots denote time derivatives. The last term in Eq.~\eqref{CLV0} is a counterterm, analogous to Eq. \eqref{CL},  which arises from the renormalization of the potential due to the coupling with the bath.

The general nature of the $F[\dot{Q}]$ coupling constitutes a set of theories, each with its own physics. We should note that, to have time-reversal symmetry, we should choose $F[\dot{Q}]$ such that it is odd in the particle velocity $\dot{Q}$. The corresponding equation of motion is now a generalized Langevin equation,
\begin{widetext}
\begin{align}  \label{protoLE}
&M\ddot{Q}+V'[Q] + F''[\dot{Q}(t)] \ddot{Q}(t) \int\limits_0^tdt'\sum\limits_k \frac{\tilde{C}_k^2}{m_k \omega_k} \sin\blockbrack{\omega_k(t-t')}F[\dot{Q}(t')]
\nonumber
+ F'[\dot{Q}(t)] \int\limits_0^tdt'\sum\limits_k \frac{\tilde{C}_k^2}{m_k} \cos\blockbrack{\omega_k(t-t')}F[\dot{Q}(t')]
\nonumber \\
& = - \xi^{(0)}(\dot{Q}(t);t)\ddot{Q}(t) - \zeta^{(0)}(\dot{Q}(t);t) + \sum\limits_k \frac{\tilde{C}_k^2}{m_k \omega_k^2} \bigg\{F'[\dot{Q}(t)]^2 + F[\dot{Q}]F''[\dot{Q}(t)]\bigg\} \ddot{Q}(t),
\end{align}
\end{widetext} for the velocity-dependent, amplitude-driven, fluctuation forces,
\begin{equation}
\xi^{(0)}(\dot{Q}(t);t)= F''[\dot{Q}(t)]f^{(0)}(t),
\end{equation}
and
\begin{equation}
\zeta^{(0)}(\dot{Q}(t);t)=F'[\dot{Q}]\frac{df^{(0)}(t) }{dt},
\end{equation}
where we have defined the bath-driven fluctuation force 
\begin{equation}
f^{(0)}(t)= \sum\limits_k \tilde{C}_k \left( q_k^{(0)}\cos(\omega_k t) + \frac{\dot{q}_k^{(0)}}{\omega_k}\sin(\omega_k t)\right).
\end{equation} {The microscopic coupling parameters $\tilde{C}_k$ are related to a phenomenological spectral function of the bath by the same relation given by Eq.~(\ref{SD}), with $J$ replaced by $\tilde{J}$ and ${C}_k$ by $\tilde{C}_k$. After partial integration and algebraic manipulations (see appendix \ref{apA}), one can rewrite  Eq.~\eqref{protoLE} in terms of the spectral function $\tilde{J}$, substituting the discrete sums over oscillators by the continuous integrals over frequency as defined in Eq.~(\ref{useJ})}. We then obtain our first general result as \begin{widetext}
 \begin{align}  \label{genresult10}
M\ddot{Q}+&V'[Q] - \int\limits_0^tdt' \int\limits_0^{\infty}
\frac{d\omega}{\pi} \frac{\tilde{J}(\omega)}{\omega}\cos\blockbrack{\omega(t-t')}
\left\{2 F''[\dot{Q}(t)] \ddot{Q}(t) F'[\dot{Q}(t')] \ddot{Q}(t')
- 2F'[\dot{Q}(t)] F[\dot{Q}(t')] \omega^2 \right\}
\nonumber \\
& = - \bar{\xi}^{(0)}(\dot{Q}(t);t)\ddot{Q}(t) - \zeta^{(0)}(\dot{Q}(t);t) + 2\int\limits_0^{\infty}\frac{d\omega}{\pi} \frac{\tilde{J}(\omega)}{\omega} F'[\dot{Q}(t)]^2 \ddot{Q}(t),
\end{align}
where
\begin{equation}
\bar{\xi}^{(0)}(\dot{Q}(t);t)=F''[\dot{Q}]\sum\limits_k \tilde{C}_k \left\{ \left[q_k(0)+\frac{\tilde{C}_k}{m_k\omega_k}F[\dot{Q}(0)]\right]\cos(\omega_kt)+\frac{\dot{q}_k}{\omega_k}\sin(\omega_kt)\right\}.
\end{equation}
\end{widetext}
 This is as far as one can go without either specifying the coupling or the spectral function. It is clear that this generalized Langevin equation is usually non-Markovian, allowing for memory dependence. This is similar to many known non-linear systems. To note, it is often possible that, even though a system has memory, it still has some memoryless subsystems \cite{Gessner}. Below, we will specify this general results under suitable assumptions about the coupling function and the properties of the bath.

\section{L\'evy statistics in Sisyphus cooling} \label{secLevy}
To gain some insight into the problem, it is convenient to analyze some limiting cases. A natural way to simplify Eq.~\eqref{genresult10} is to assume that the second derivative of the coupling is very small compared to the other terms, i.e.
$F''[\dot{Q}]\sim 0.$ This choice will be justified below in the context of subrecoil cooling.
The general Eq.~\eqref{genresult10} is then considerably simplified and we obtain
 \begin{eqnarray}
 \label{smallddLang}  \nonumber
&&M\ddot{Q}+2 \int\limits_0^t dt'\int\limits_0^{\infty}\hspace{-0.1cm}\frac{d\omega}{\pi} \omega \tilde{J}(\omega)\cos\blockbrack{\omega(t-t')}F'[\dot{Q}(t)]F[\dot{Q}(t')] \\ \nonumber &&+ V'[Q] \approx   - \zeta^{(0)}(\dot{Q}(t);t) + 2\int\limits_0^{\infty}\frac{d\omega}{\pi} \frac{\tilde{J}(\omega)}{\omega} F'[\dot{Q}(t)]^2 \ddot{Q}(t). \\
 \end{eqnarray}
This is one of the main results of this work, The generic kernel in Eq. \eqref{smallddLang} can describe both Markovian processes like Brownian and Lévy motion as well as non-Markovian processes like fractional Brownian motion and fractional Lévy motion (cf. Ref. \cite{Verstraten}).

\begin{figure}[bth] 
\centering
\includegraphics[width=0.5\textwidth]{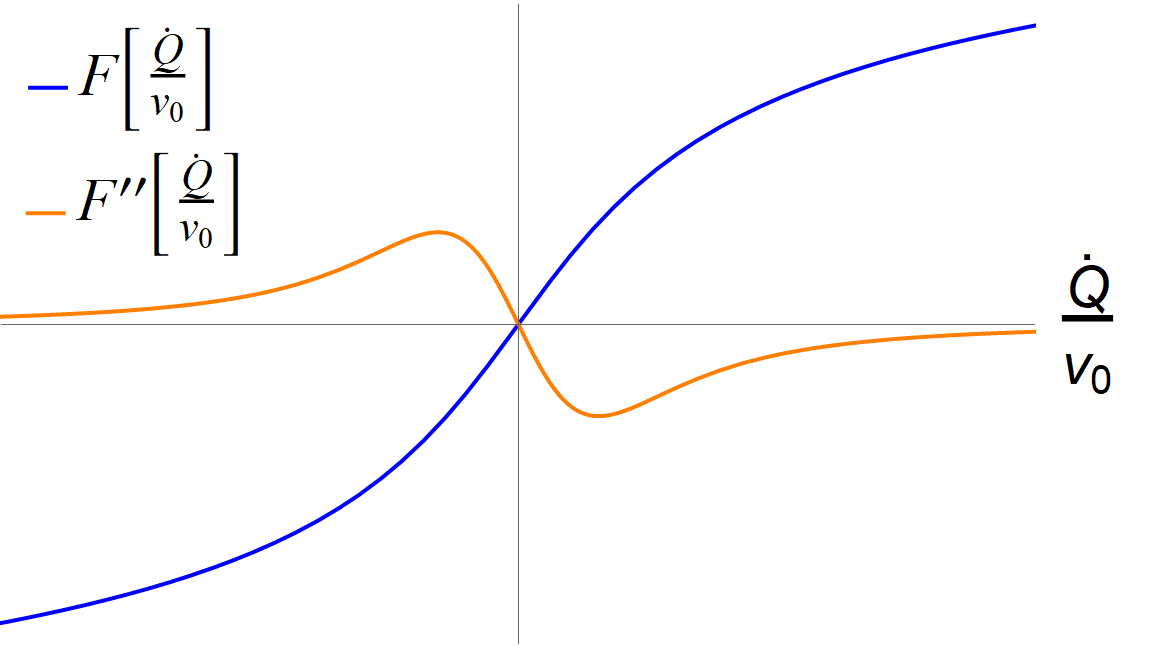}
\caption{The coupling function of Eq.~\eqref{Zollercoup0}. One can see that in both the high- and the low-velocity limits, $F''[\dot{Q}]$ is consistently negligible. }
\label{fig:11}
\end{figure}

Now, we can establish a connection with L\'evy behavior in ultracold atoms and try to reproduce the friction force from Ref. \cite{Marksteiner}, $ f[\dot{Q} ] = \dot{Q} / [1+(\dot{Q}/v_0)^2]$, also stated above by Eq. \eqref{Marksteiner} in terms of (canonical) momentum $p$. The constant $v_0$ contains the mass of the particle and the specifics of the physical set-up, e.g. the experimental properties of the laser during \textit{Sisyphus} cooling \cite{Lutz}. During \textit{Sisyphus} cooling, individual atoms in a cooling cloud exhibit L\'evy statistics as a result of the recoil caused by spontaneous photon emission. It turns out that in this context  an appropriate choice for the coupling is
\begin{equation} \label{Zollercoup0}
F[\dot{Q}]= \text{sgn}\left(\frac{\dot{Q}}{v_0}\right) v_0  \sqrt{\log\left[1+\left(\dot{Q}/ v_0\right)^2\right]},
\end{equation}
where the sign function $\text{sgn} (x)$ is included to ensure both a smooth transition of the first derivative at $\dot{Q}=0$ and a unique value for $F[\dot{Q}]$ for every value of $\dot{Q}$ (see Fig.~\ref{fig:11} for a sketch of the coupling).

To connect this to earlier work on L\'evy statistics in optical lattices, we must calculate the derivative of this coupling term.  We then obtain for the friction force term in the Langevin equation (see appendix \ref{apB})
\begin{equation} \label{FZoller}
F[\dot{Q}]F'[\dot{Q}]= \frac{\dot{Q}}{1+\left(\dot{Q} / v_0 \right)^2},
\end{equation}
which coincides with Eq. \eqref{Marksteiner}, which is the same as Eq. (37) of Ref. \cite{Marksteiner} and Eq. (5) of Ref. \cite{Kessler}.
This non-linear friction force is called the `cooling force', as it acts to bring back the momentum of the atom to  $p=0$: the minimum energy state. As Ref. \cite{Kessler}, we note that similar friction forces with a $1/p$ momentum dependence, like Eq. \eqref{FZoller}, are common in other fields of research also, for example for an atomic tip on a surface at cryogenic temperatures, Eq. (3) in Ref. \cite{Barel}.
 
It can be promptly verified that the second derivative of the force $F''[\dot{Q}]$  is consistently negligible in both the  high-velocity limit $\dot{Q} \gg v_0$ and the  low-velocity limit $\dot{Q} \ll v_0$, as can be seen in Fig.~\ref{fig:11}. Thus, the approximation is justified. {Considering the particular coupling give by Eq.~(\ref{Zollercoup0}), for dimensional reasons we have to choose $ \tilde{C}_k = C_k/\omega_k$. From the definition in Eq.~(\ref{useJ}), it is easy to see that this implies the relation $\tilde{J}(\omega) = J(\omega)/\omega^2$. Now, depending on the choice of the spectral function $J(\omega)$, the processes will be Markovian (ohmic) or non-Markovian (subohmic and superohmic). Choosing an ohmic bath, we  get a similar Langevin equation as shown in \cite{Kessler} (see appendix \ref{apB} for further details in how to get the memory-free Kernel for the drift coefficient in Eq. (\ref{smallddLang})).}

\section{The linear-velocity approach for a superohmic bath}

We proceed by analyzing the limit $\dot{Q} \ll v_0$ in more detail. In this case, the denominator of Eq.~\eqref{FZoller} reduces to unity and the force term is simply linear in the velocity $\dot{Q}$. {We now assume a superohmic (cubic) spectral function, \begin{equation} \label{Jnietechtohm0}
{J}(\omega) =
\left\{
	\begin{array}{ll}
		\lambda \omega^3  & \mbox{if } \omega < \Omega \\
		0	  & \mbox{if } \omega > \Omega
	\end{array}
\right.,
\end{equation}}
 where $\Omega$ corresponds to a high-frequency cut-off and the diffusion coefficient is denoted by $\lambda$. The equation of motion then reduces to (see appendix \ref{apC}) \begin{align}  \label{jerkEOM0}
M\ddot{Q}(t)+V'[Q] -2 \lambda  \dddot{Q}(t)   =   - \bar{\zeta}^{(0)}(t),
 \end{align}
 
The spectral properties of the bath are in analogy with \cite{Barone}, where the Caldeira-Leggett model for cubic baths was used to describe the dissipation of an electron interacting with its own
radiation field. Classically, such an electron is described by the Abraham-Lorentz equation,  
\begin{equation} \label{ALE}
-\frac{2e^2}{3c^3} \dddot{Q}+M^*\ddot{Q}+V'[Q] = f^{(0)},
\end{equation}
with the fluctuation force given by \begin{equation} \label{forcesum}
f^{(0)}\hspace{-0.1cm}=\sum\limits_k C_k f^{(0)}_k\hspace{-0.1cm}= \sum\limits_k C_k\hspace{-0.1cm} \left( q_k^{(0)}\cos(\omega_k t) + \frac{\dot{q}_k^{(0)}}{\omega_k}\sin(\omega_k t)\hspace{-0.1cm}\right),
\end{equation} for $\omega_k=c k$, for wavenumber $k$ and speed of light $c$---the dispersion relation for the photon. The Abraham-Lorentz equation has important problems of its own. For example, pre-acceleration, where the acceleration at time $t$ depends on the value of the radiation force at earlier times $t'$, which is, therefore, a memory effect.

Would we have chosen $J(\omega)$ to be ohmic in this low-velocity limit approximation, then we would have retrieved the usual Markovian Brownian motion. Notice that in the opposite limit $\dot{Q} \gg v_0$ we will obtain either Lévy motion in the Markovian limit or fractional Lévy motion when the kernel retains memory \cite{Verstraten}. In this low-velocity regime, the L\'evy behaviour is not visible. Yet, the model specified by Eqs.~\eqref{CLV0} and \eqref{Zollercoup0} provide a  connection between L\'evy behaviour and an equation analogous to the Abraham-Lorentz Eq.~\eqref{ALE}. This is not unexpected for a Caldeira-Leggett model: Barone and Caldeira  reproduced the Abraham-Lorentz equation from first principles for a superohmic bath. They started with the Lagrangian
\begin{equation} \label{BaroneLag}
\mathcal{L}=\mathcal{L}_S+\mathcal{L}_B+\mathcal{L}_{I}
\end{equation}
 where the system Lagrangian is generic, the bath consists of the electrodynamic field associated with the self-field of the electron
\begin{equation}
\mathcal{L}_B=\frac{1}{8\pi}\int d\mathbf{x} \blockbrack{\frac{1}{c}\frac{\partial \mathbf{A}}{\partial t} + \mathbf{\nabla}\phi}^2-(\mathbf{\nabla}\times\mathbf{A})^2,
\end{equation}
and the system-bath interaction is given by
\begin{equation}
\mathcal{L}_I=\int d\mathbf{x} \blockbrack {\frac{1}{c}\mathbf{J}\cdot \mathbf{A}-\rho \phi },
\end{equation}
 where $\mathbf{A}$ and $\phi$ are the vector and scalar potentials, and $\rho$ and $\mathbf{J}$ the charge and current densities. Note that there is no counterterm explicitly assumed in this model.

 By changing to the Hamiltonian formalism, writing the expressions in terms of conjugated momenta of the coordinates $Q$ and $\mathbf{A}$, and assuming a superohmic cubic spectral density of the form \eqref{Jnietechtohm0}, they were able to derive the equation of motion
\begin{equation} \label{ALE2}
- \lambda \dddot{Q} + M^*\ddot{Q}+ V'[Q]=f(t),
\end{equation}
for
\begin{equation}
\lambda=\frac{2e^2}{3c^3},
\end{equation}
and with the renormalized mass  given by
\begin{equation} \label{effmass}
M^*=M+\frac{2 \lambda \Omega}{\pi},
\end{equation}
 for the bare electron charge $e$, bare electron mass $M$, and speed of light $c$. Here, we also see that the dissipation term is proportional to the jerk, which is a direct consequence of the low-frequency behavior of the spectral distribution in a cavity \cite{Caldeira}. The renormalized mass has the usual interpretation of a mass renormalization in terms of the screening of the bare electron mass by the one-loop-and-higher Feynman diagrams of the electron propagator---which is simply what the electron self-interaction is in quantum-electrodynamical language.

The merit of having written down the Lagrangian \eqref{BaroneLag} in terms of the fundamental constants of electrodynamics is that Barone and Caldeira can now estimate how large the cut-off-dependent terms in the equations of motion are. The cut-off itself is given by
\begin{equation}
\Omega=\frac{2\pi c}{r_0},
\end{equation}
where $r_0$ is the characteristic electronic dimension. It is the classical electron radius in the classical limit, while it is the deBroglie wavelength of the electron in the quantum case.
 The effective mass depends on the cut-off $\Omega$, but since the numerical values in Eq.~\eqref{effmass} are known, we can conclude $M^* \sim M$.

Coming back to our result Eq.~\eqref{jerkEOM0}, we see that it has exactly the same form as the Abraham-Lorentz equation. 
This verifies the consistency of our general result in the low-velocity limit and connects our findings of super-diffusion
to other semiclassical models \cite{Kessler, Barkai, Dechant}. However, whereas Caldeira and Barone assumed a specific electrodynamical model to derive the Eq.~\eqref{ALE2}, for our result we did not assume any fundamental physics. Hence, in our case, this could be seen as a self-interaction of a general physical system---as long as the coupling is not too strong.

Furthermore, Barone and Caldeira note that the ``counterterm is not imposed as in the formulation of the quantum Brownian motion [...]'' \cite{Barone}. The reason is that the counterterm appears naturally from the microscopic set-up of the model. This remark is important, since in our model we have also not included a counterterm as in Eq.~\eqref{counterterm} for quantum Brownian motion.

\section{Quantizing the generalized model}

The next aim is  to construct a quantum version of the extended Caldeira-Leggett Lagrangian in terms of the generic velocity coupling that we have introduced. We will use the path-integral formalism to calculate the propagator for the Lagrangian~\eqref{CLV0}. Since we are only interested in the stochastic particle, we use the reduced density operator strategy and integrate out the bath from the full density operator to obtain an effective dynamics for the system of interest. As a first step, we write down the Euclidean action corresponding to the Lagrangian~\eqref{CLV0}. Then, we evaluate the influence functional $\mathcal{F}$ in Euclidean time, after performing a Wick rotation and find (see appendix \ref{apD}) \begin{align} \label{IFtot}
\mathcal{F}
=  \exp \left[\sum\limits_{k=1}^N A_k
\int\limits_0^{\hbar \beta}d\tau \int\limits_0^{\tau}d\sigma   \mathcal{G}_k(\tau - \sigma) F[\dot{Q}(\tau)]F[\dot{Q}(\sigma)] \right]
\end{align}
where  $ A_k =  \tilde{C}_k^2/ 2\hbar m_k \omega_k $ and the kernel
$\mathcal{G}_k(\hbar \beta - \tau)= \mathcal{G}_k(\tau)\cosh\left[\omega_k\left(\hbar \beta/2 -\tau\right) \right]/ \sinh\left(\hbar \beta \omega_k/ 2 \right)$ is a periodic function.

The bath particles have now been successfully integrated out. Their presence is seen through the double integral and the occurrence of the forces $F$. This term contains non-local interactions, which express the influence of the bath on the dynamics of the stochastic particle.

We are now in a position to write the reduced density operator of the system,
\begin{equation}
\hat{\rho}_R(Q,Q')= 
\int\limits_{Q'}^Q \mathcal{D} Q \exp \left(-\frac{S^{E}_{\text{eff}}[Q]}{\hbar} \right)
\end{equation}
where the effective action has the form
\begin{widetext}
\begin{equation} \label{effactCT}
S^{E}_{\text{eff}}=\int\limits_0^{\hbar \beta}d\tau \left(\frac{1}{2}M\dot{Q}^2+V(Q)-\sum\limits_{k=1}^N\frac{\tilde{C}_k^2}{2m_k\omega_k^2}F[\dot{Q}]^2\right) +\sum\limits_{k=1}^N \frac{\tilde{C}_k^2}{2m_k \omega_k}
\int\limits_0^{\hbar \beta}d\tau \int\limits_0^{\tau}d\sigma   \mathcal{G}_k(\tau - \sigma) F[\dot{Q}(\tau)]F[\dot{Q}(\sigma)].
\end{equation}
Using known manipulations \cite{Hibbs,Vernon}, we split the domains of integration and use the counterterm to complete the square, such that the effective action becomes $S^{E}_{\text{eff}}= S^{E}_S + S^{E}_D$, with
 \begin{equation} \label{geneffact}
S^{E}_S = \int\limits_0^{\hbar \beta}d\tau \left\{\frac{1}{2}M\dot{Q}^2+V(Q)\right\}, \qquad
S^{E}_D = \frac{1}{4\pi}\int\limits_0^{\hbar \beta}d\tau \int\limits_{-\infty}^{\infty}d\sigma \int\limits_0^{\infty}d\omega \tilde{J}(\omega)e^{-\omega|\tau-\sigma|}\blockbrack{F[\dot{Q}(\tau)]-F[\dot{Q}(\sigma)]}^2.
\end{equation}
\end{widetext}
This is the second main result of this work. For `ohmic' dissipation, $\tilde{J}(\omega)=\lambda \omega$, which corresponds to superohmic dissipation  $J(\omega)$,
after performing the integral over $\omega$ for $\Omega \rightarrow \infty$, one finds the final result for the dissipation term in the effective Euclidean action
\begin{equation} \label{ohmiceffact}
S^E_D[Q(\tau)]=\frac{\lambda}{4\pi}\int\limits_0^{\hbar \beta} d\tau\int\limits_{-\infty}^{\infty} d\sigma \frac{\left\{F[\dot{Q}(\tau)]-F[\dot{Q}(\sigma)]\right\}^2}{|\tau-\sigma|^2}.
\end{equation}
This term is entirely induced by the bath. It is analogous to the `Caldeira-Leggett'-kernel, but it is more general. If we would replace the generic coupling by Eq. \eqref{Zollercoup0}, the quantum version of the model that gives rise to Lévy behaviour in the semi-classical limit result for $\dot{Q} \gg v_0$.

In our case, the coupling is not linear, but generic, and in terms of velocity. If we would have started out with a coordinate-coordinate coupling and counterterm such as in the original Caldeira-Leggett model, the result would be exactly Eq.~\eqref{geneffact} with $F$ replaced by $Q$. In that case, it could be promptly understood that the Caldeira-Leggett kernel describes friction: for $|\tau-\sigma| \sim 0$, the term diverges unless the relative position of the Brownian particle at different times, $Q(\tau)-Q(\sigma)$ goes to zero, which is physically interpreted as the tendency to oppose motion.

Let us now investigate the generalized equation Eq.~\eqref{ohmiceffact} for $F[\dot{Q}]$ given by Eq.~\eqref{Zollercoup0} in the low-velocity limit, $\dot{Q} \ll v_0$. We then find that the integrand of the dissipation term in the action reduces to  $(\lambda/4\pi) \blockbrack{\dot{Q}(\tau)-\dot{Q}(\sigma)}^2/|\tau-\sigma|^2$. This term is non-local, since it exclusively depends on the relative velocity of the Brownian system at different times. The interpretation of this \textit{modified} \textit{term} can be made in a similar way as for the original Caldeira-Leggett model. For successive times close together, i.e. $|\tau-\sigma| \sim 0$, this term diverges unless the relative velocity of the Brownian system at different times, $\dot{Q}(\tau)-\dot{Q}(\sigma)$, also tends to zero. Hence, it is energetically better for the particle to resist acceleration, and we can interpret this bath-induced effect as that of \textit{inertia} on top of the mass $M$ in the local part of the action.

\section{Conclusion and Outlook}

The goal of this work was twofold. First, to construct a closed Lagrangian model that could reproduce L\'evy motion. The motivation for this was largely to account for a description of L\'evy statistics during \textit{Sisyphus} cooling. Second, to construct a quantum version of such a system via path integral quantization. To reach these goals, we modified the Caldeira-Leggett model to general velocity-dependent coupling. With this choice, we left other possibilities behind, e.g. trying to construct a stochastic Schr\"odinger equation in the style of Kostin \cite{Kostin} and Nelson \cite{Nelson}, using a modified quantization scheme \cite{Dekker}, or entering the Lindblad formalism \cite{Lindblad}. One advantage of our approach is the ability to describe non-Markovian systems, which is absent from the Lindblad formalism.

Through Lagrangian~\eqref{CLV0}, we modified the interaction between the system and the bath. The resulting \textit{modified} generalized Langevin equation is given by Eq.~\eqref{genresult10}. 

For the occurrence of L\'evy walks in ultracold-atoms experiments like \textit{Sisyphus} cooling, we were able to reproduce the friction force from Ref.  \cite{Marksteiner}, which gives rise to L\'evy statistics for $\dot{Q} \gg v_0$. Here, individual atoms in a cooling cloud exhibit L\'evy walks as a result of the recoil caused by spontaneous photon emission. It turns out that the choice of Eq.~\eqref{Zollercoup0} yields exactly this expression. We have also shown that the approximation $F'' \sim 0$ holds in the low- and high-velocity limits,  but further research should be undertaken to interpolate in between these extremes, where the approximation fails and non-Markovian memory effects start playing a larger role. This will involve a numerical investigation. 

Under the approximation $F''[\dot{Q}] \sim 0$, we have worked out a concrete realization of the coupling function $F[\dot{Q}]$. For the linear-velocity case, obtained in the low-velocity limit, we derive the equation of motion Eq.~\eqref{jerkEOM0}. Without assuming any fundamental physical picture, this equation bears exact resemblance to the Abraham-Lorentz equation of a self-interacting electron. Adapting the counterterm accordingly, we do not find a mass renormalization because the new counterterm adds precisely that amount of mass to the kinetic part of the system of interest.

It is important to bear in mind that we found this under the assumption that the spectral function $\tilde{J}$ is linear in the frequency, and that this corresponds to an ohmic bath in terms of coupling constants $\tilde{C}_k$, but that for coupling constants $C_k$, the bath is cubic,  $J \propto \omega^3$, and hence superohmic. Therefore, the term `ohmic bath' is ambiguous when one does not explicitly specify the dimension of the coupling constants.

Finally, we have discussed the path-integral formulation of quantum mechanics and used the Feynman-Vernon influence-functional method in Euclidean time to integrate out the bath particles and derive an effective action for our modified Lagrangian~\eqref{CLV0}. The result, for an ohmic bath in terms of coupling constants $\tilde{C}_k$, is given by Eq.~\eqref{ohmiceffact}. It has the same form as the original Caldeira-Leggett term.

This final result should be seen as a scaffolding for further research: it is a collection of different models, where each model is specified by a particular coupling. Analogous to the interpretation of friction in the original Caldeira-Leggett term \cite{CaldLegg}, the modified term in the low-velocity limit is that of inertia, resisting acceleration. The classical equation of motion corresponding to this is the Abraham-Lorentz equation~\eqref{jerkEOM0}.

In addition to the above-mentioned numerical work to be done for several coupling functions, one can also look at the decoherence properties of the L\'evy-walk coupling~\eqref{Zollercoup0} and/or the linear-velocity case. For further calculation, one can specify a particular form for the external potential $V(Q)$. In that case, one can, in principle, determine the effective dynamics of the system of interest by performing the path integral over all paths $Q(\tau)$.


\begin{thebibliography}{1}

\bibitem{CaldLegg} A. O. Caldeira  and  A. J. Leggett,
Physica \textbf{121A}, 587 (1983).

\bibitem{Caldeira}  A. O. Caldeira, \textit{An Introduction to Macroscopic Quantum Phenomena and Quantum Dissipation} (Cambridge University Press, New York, 2014).

\bibitem{CaldLegg2}  A. O. Caldeira and A. J. Leggett,
Annals of Physics \textbf{149}, 374  (1983).

\bibitem{CaldLegg3}  A. O. Caldeira and A. J. Leggett,
Phys. Rev. A \textbf{31}, 1059 (1985).

\bibitem{Zaburdaev} V. Zaburdaev, S. Denisov, J. Klafter, Rev. Mod. Phys. \textbf{87}, 483 (2015).

\bibitem{West} M. F. Schlesinger, J. Klafter, M. F. West. 
Physica A: Statistical Mechanics and its Applications \textbf{140}, 212  (1986).

\bibitem{Rice} J. T. Bendler, J. Statist. Phys. \textbf{36}, 625 (1984).

\bibitem{Barthelemy} P.  Barthelemy, J. Bertolotti and  D. S. Wiersma, Nature \textbf{453}, 495 (2008).

\bibitem{Bardou1} F. Bardou, J.-P. Bouchaud, A. Aspect, C. Cohen-Tannoudji,
 \textit{L\'evy Statistics and Laser Cooling} (Cambridge University Press, Cambridge, 2002).

\bibitem{Kravtsov} V. E. Kravtsov, O. M. Yevtushenko, P. Snajberk, and E. Cuevas,
Phys. Rev. E \textbf{86}, 021136 (2012).

\bibitem{Kempkes} S. N. Kempkes, M. R. Slot, S. E. Freeney, S. J. M. Zevenjuizen. D. Vanmaekelbergh, I. Swart, and C. Morais Smith, Nature Phys. \textbf{15}, 127 (2019).

\bibitem{Lutz} Lutz, E., Renzoni, Nature Phys. \textbf{9}, 615 (2013).

\bibitem{Afek} Afek, Gadi and Coslovsky, Phys. Rev. Lett. \textbf{119}, 060602 (2017).

\bibitem{Barone}  P. M. V. B. Barone and A. O. Caldeira,
Phys. Rev. A \textbf{43}, 57 (1991).

\bibitem{Levy} P. L\'evy, \textit{Th\'eorie de l'addition des variables al\'eatoires}. Paris: Gauthier-Villars (1937).

\bibitem{Klafter} J. Klafter, M. F. Shlesinger, G. Zumofen, Physics Today \textbf{49}, 33 (1996).
 
\bibitem{Katori} H. Katori, S. Schlipf, and H. Walther, Phys. Rev. Lett. \textbf{79}, 2221 (1997).

\bibitem{Bardou2} F. Bardou, J. P. Bouchaud, O. Emile, A. Aspect, and C. Cohen-Tannoudji, Phys. Rev. Lett. \textbf{72}, 203 (1994).
  
\bibitem{Sagi} Y. Sagi, M. Brook, I. Almog, and N. Davidson, Phys.Rev. Lett. \textbf{108}, 093002 (2012).

\bibitem{Barkai} E. Barkai, E. Aghion, and D. A. Kessler, Phys. Rev. X \textbf{4}, 021036 (2014).

\bibitem{Peter} H. J. Metcalf, P. van der Straten, \textit{Laser Cooling and Trapping}. New York: Springer-Verlag (1999).
   
\bibitem{Marksteiner} S. Marksteiner, K. Ellinger, P. Zoller, 
Phys. Rev. A \textbf{53}, 3409 (1996). 

\bibitem{Kessler} D. A. Kessler, and E. Barkai, Phys. Rev. Lett. \textbf{108}, 230602 (2012).

\bibitem{Weiss} U. Weiss, \textit{Quantum Dissipative Systems}(World Scientific Publishing, Singapore. 1992), 2nd edition.

\bibitem{Gessner}  M. Gessner, \textit{Dynamics and Characterization of Composite Quantum Systems} (Springer International Publishing, Cham, 2017).

\bibitem{Verstraten} R. C. Verstraten, R. F. Ozela, C. Morais Smith (2020). ``Time Glass: A Fractional Calculus Approach." Prepint. Preprint available here: \href{https://arxiv.org/abs/2006.08786}{2006.08786 [cond-mat.stat-mech]}.

\bibitem{Barel} I. Barel, M. Urbakh, L. Jansen, and A. Schirmeisen, Phys. Rev. Lett. \textbf{104}, 066104 (2010).

\bibitem{Dechant} A. Dechant, D. A.  Kessler and E. Barkai, Phys. Rev. Lett. \textbf{115}, 173006 (2015).

\bibitem{Hibbs} R. P. Feynman, A. R. Hibbs, \textit{Quantum Mechanics and Path Integrals}(McGraw-Hill Companies, New York, 1965).

\bibitem{Vernon}  R. P. Feynman and L. Vernon, Jr.
Ann. Phys. (NY) \textbf{24}, 118 (1963).

\bibitem{Kostin} M. D. Kostin, .
J. Chem. Phys. \textbf{57}, 3589 (1972).

\bibitem{Nelson} E. Nelson,
Phys. Rev. \textbf{150}, 1079 (1966).

\bibitem{Dekker}H.  Dekker,
Phys. Rev. A \textbf{16}, 2116 (1977).

\bibitem{Lindblad} G. Lindblad,
Comm. Math. Phys. \textbf{48}, 119 (1976).
\end{thebibliography}

\section{Acknowledgments}
We are grateful to Peter Zoller for fruitful discussions. M. A. C.  acknowledges financial support from CNPq (Brazil) through the Brazilian government project `Science Without Borders' [Process number: 202091/2015-5].

\appendix

\section{Generalized Caldeira-Leggett Model} \label{apA}

We consider the Lagrangian \begin{eqnarray} \label{CLV}
\mathcal{L}&=& \nonumber
\frac{1}{2}M\dot{Q}^2-V(Q)
+
\frac{1}{2}\sum\limits_{k=1}^Nm_k(\dot{q}^2_k-\omega_k^2 q_k^2) \\
&+&
F[\dot{Q}]\sum\limits_{k=1}^N \tilde{C}_kq_k - \sum\limits_{k=1}^N\frac{\tilde{C}_k^2}{2m_k\omega_k^2}F^2[\dot{Q}].
\end{eqnarray} \\ 

\noindent From Eq.~(\ref{CLV}) we derive the system equation of motion \begin{eqnarray}  \label{seom3} \nonumber
&&M\ddot{Q}+V'(Q) + F''[\dot{Q}] \ddot{Q} \sum\limits_k \tilde{C}_k q_k  + F'[\dot{Q}] \sum\limits_k \tilde{C}_k \dot{q}_k \\ &&= \sum\limits_k \frac{\tilde{C}_k^2}{m_k \omega_k^2} \blockbrack{F'[\dot{Q}]^2 + F[\dot{Q}]F''[\dot{Q}]} \ddot{Q},
\end{eqnarray} where the right-hand side is entirely due to the counterterm. The bath equation of motion is a driven harmonic oscillator equation,
\begin{align} \label{bath3}
m_i\ddot{q}_i+m_i\omega_i^2 q_i&= \tilde{C}_i F(\dot{Q}),
\end{align}
with solution
 \begin{equation} \label{sbath3}
 q_i(t)=f_i^{(0)}(t) +
 \frac{\tilde{C}_i}{m_i \omega_i}\int\limits_0^tdt'\sin[\omega_i(t-t')]F[\dot{Q}(t')],
 \end{equation}
 and its time-derivative
  \begin{equation} \label{dsbath3}
 \dot{q}_i(t)=\frac{df_i^{(0)}(t) }{dt}
 + \frac{\tilde{C}_i}{m_i}\int\limits_0^tdt'\cos[\omega_i(t-t')]F[\dot{Q}(t')].
 \end{equation} 
Let us now plug Eqs. \eqref{sbath3} and \eqref{dsbath3} into the system equation \eqref{seom3}, such that we obtain a differential equation analogous to the generalized Langevin equation. We obtain then
\begin{widetext}
 \begin{align}  \label{protoLEa}
M\ddot{Q}+&V'[Q] + F''[\dot{Q}(t)] \ddot{Q}(t) \int\limits_0^tdt'\sum\limits_k \frac{\tilde{C}_k^2}{m_k \omega_k} \sin\blockbrack{\omega_k(t-t')}F[\dot{Q}(t')]
+ F'[\dot{Q}(t)] \int\limits_0^tdt'\sum\limits_k \frac{\tilde{C}_k^2}{m_k} \cos\blockbrack{\omega_k(t-t')}F[\dot{Q}(t')]
\nonumber \\
& = - \xi^{(0)}(\dot{Q}(t);t)\ddot{Q}(t) - \zeta^{(0)}(\dot{Q}(t);t) + \sum\limits_k \frac{\tilde{C}_k^2}{m_k \omega_k^2} \blockbrack{F'[\dot{Q}(t)]^2 + F[\dot{Q}]F''[\dot{Q}(t)]} \ddot{Q}(t),
\end{align}
\end{widetext}
for the velocity-dependent and amplitude-driven fluctuation forces,
$\xi^{(0)}(\dot{Q}(t);t)= F''[\dot{Q}(t)]f^{(0)}(t)$  and $\zeta^{(0)}(\dot{Q}(t);t)=F'[\dot{Q}]df^{(0)}(t)/dt$.  Next, we integrate by parts the third term on the left-hand side of Eq.~\eqref{protoLEa}, 
\begin{widetext}
which then becomes
\begin{align}
 -F''[\dot{Q}(t)] \ddot{Q}(t) & \int\limits_0^tdt'\sum\limits_k \frac{\tilde{C}_k^2}{m_k \omega_k^2} \cos\blockbrack{\omega_k(t-t')}F'[\dot{Q}(t')]\ddot{Q}(t')
 \nonumber \\
 &+\sum\limits_k \frac{\tilde{C}_k^2}{m_k \omega_k^2}F''[\dot{Q}(t)] \ddot{Q}(t)F[\dot{Q}(t)]-\sum\limits_k \frac{\tilde{C}_k^2}{m_k \omega_k^2}F''[\dot{Q}(t)] \ddot{Q}(t)F[\dot{Q}(0)] \cos(\omega_k t).
\end{align}
\end{widetext}
Now, the first boundary term will cancel one of the counterterms [the last term on the right-hand side of Eq.~\eqref{protoLEa}]. The other boundary term depends on the initial value $\dot{Q}(0)$, which can be incorporated into $\xi^{(0)}(Q(t);t)$ as follows
\begin{widetext}
\begin{eqnarray} \nonumber
\bar{\xi}^{(0)}(Q\dot{}(t);t)=F''[\dot{Q}]\sum\limits_k \tilde{C}_k\curlbrack{\blockbrack{q_k(0)  +\frac{\tilde{C}_k}{m_k\omega_k}F[\dot{Q}(0)]}\cos(\omega_kt)+\frac{\dot{q}_k}{\omega_k}\sin(\omega_kt)}.
\end{eqnarray}
Hence, we can rewrite the  Langevin equation as
\begin{align}
M\ddot{Q}+ V'[Q] - F''[\dot{Q}(t)] \ddot{Q}(t) &\int\limits_0^tdt'\sum\limits_k \frac{\tilde{C}_k^2}{m_k \omega_k^2} \cos\blockbrack{\omega_k(t-t')}F'[\dot{Q}(t')] \ddot{Q}(t') + F'[\dot{Q}(t)] \int\limits_0^tdt'\sum\limits_k \frac{\tilde{C}_k^2}{m_k} \cos\blockbrack{\omega_k(t-t')}F[\dot{Q}(t')]
\nonumber \\
& = - \bar{\xi}^{(0)}(\dot{Q}(t);t)\ddot{Q}(t) - \zeta^{(0)}(\dot{Q}(t);t) + \sum\limits_k \frac{\tilde{C}_k^2}{m_k \omega_k^2} F'[\dot{Q}(t)]^2 \ddot{Q}(t).
\end{align}
\end{widetext}

\section{L\'evy walks: A non-polynomial coupling} \label{apB}
In Section~\ref{secLevy} we eliminate direct memory terms assuming that the second derivative of the coupling is very small compared to the other terms in the Langevin equation \eqref{genresult10}, i.e.
\begin{equation} \label{smalld0}
F''[\dot{Q}]\sim 0.
\end{equation}
This simplifies the generalized Langevin equation \eqref{genresult10} considerably. Note that Eq.~\eqref{smalld0} does not necessarily imply that we must choose the linear-velocity case, for which it is trivially satisfied. This is where we establish a connection with L\'evy statistics in ultracold atoms. As explained in the main text, to reproduce Marksteiner, Ellinger and Zoller's friction force that we saw in Section \ref{LaserCool}, we choose
\begin{eqnarray} \label{Zollercoup}  
F[\dot{Q}]= \text{sgn}\left(\frac{\dot{Q}}{v_0}\right) v_0  \sqrt{\log\blockbrack{1+\left(\frac{\dot{Q}}{v_0}\right)^2}},
\end{eqnarray} (see \textsf{Figure} \ref{fig:11} in the main text).

To see the connection to the force \eqref{Marksteiner}, we must calculate the derivative of this coupling term. It is
\begin{eqnarray} \nonumber
F'[\dot{Q}]
&=&\text{sgn}\left(\frac{\dot{Q}}{v_0}\right)v_0  \frac{\dot{Q}}{ \sqrt{\log\blockbrack{1+\left(\frac{\dot{Q}}{v_0}\right)^2}}\blockbrack{1+\left(\frac{\dot{Q}}{v_0}\right)^2}} \\ 
&=&    \frac{\dot{Q}}{F[\dot{Q}]\blockbrack{1+\left(\frac{\dot{Q}}{v_0}\right)^2}}.
\end{eqnarray}
 We then obtain for the force term in the Langevin equation \eqref{smallddLang}
\begin{equation}
F[\dot{Q}]F'[\dot{Q}]= \frac{\dot{Q}}{1+\left(\frac{\dot{Q}}{v_0}\right)^2},
\end{equation}
which coincides with Eq.~\eqref{Marksteiner}. \\

{For the drift coefficient (second term in the left side of Eq.(\ref{smallddLang})), considering the relation $\tilde{J}(\omega)= J(\omega)/\omega^2$, for an ohmic bath $J(\omega) = \eta \omega$ we will have
\begin{eqnarray}
 \label{drift coefficient}  \nonumber
&&2 \int\limits_0^t dt'\int\limits_0^{\infty}\frac{d\omega}{\pi}  \frac{J(\omega)}{\omega}\cos\blockbrack{\omega(t-t')}F'[\dot{Q}(t)]F[\dot{Q}(t')] \\ \nonumber &&= 2\eta \int\limits_0^t dt'\int\limits_0^{\infty}\frac{d\omega}{\pi} \cos\blockbrack{\omega(t-t')}F'[\dot{Q}(t)]F[\dot{Q}(t')] \\  &&= 2\eta F'[\dot{Q}(t)]F[\dot{Q}(t)] = \frac{2\eta \dot{Q}(t)}{1+\left(\frac{\dot{Q}(t)}{v_0}\right)^2},  
 \end{eqnarray} where the delta distribution $\delta(t-t')$ resulting from the frequency integral give us the local-time drift coefficient.}

\section{Polynomial coupling and the linear-velocity approach for a superohmic bath} \label{apC}

Now we will discuss the generalized Langevin equation for a system of linear-velocity coupling 
\begin{equation} \label{lincoup}
\tilde{C}_k F[\dot{Q}]= \frac{{C}_k}{\omega_k}\dot{Q},
\end{equation}
where we related the general-velocity coupling constant (where we had a `tilde') with that from the original Caldeira-Leggett model \cite{Weiss} for the linear-velocity case we will use now. 

For the coupling \eqref{lincoup}, the generalized Langevin equation \eqref{genresult10} simplifies considerably, since the first derivatives of the coupling become unity, whereas the second derivatives vanish. The result is
 \begin{eqnarray}  \label{lvcLang} \nonumber
 M\ddot{Q}(t)&+&V'[Q] + 2  \int\limits_0^tdt'  \int\limits_0^{\infty}\frac{d\omega}{\pi} \frac{{J}(\omega)}{\omega}\cos\blockbrack{\omega(t-t')}\dot{Q}(t')\\ 
& =&   - \zeta^{(0)}(t) + 2\int\limits_0^{\infty}\frac{d\omega}{\pi} \frac{{J}(\omega)}{\omega^3}  \ddot{Q}(t),
\end{eqnarray}
for the force
\begin{equation} \label{Zeta}
\zeta^{(0)}=\frac{d}{dt} \sum\limits_k  \frac{{C}_k}{\omega_k} \left( q_k^{(0)} \cos(\omega_k t)+\frac{\dot{q}_k^{(0)}}{\omega_k} \sin{(\omega_k t)} \right).
\end{equation}

\noindent To show that $ \zeta^{(0)}(t)$ is really a fluctuation force, one needs to calculate the average of the fluctuation force  and its two-point correlation function. For that we use the initial conditions for an equilibrium bath at temperature $T$, which, through the equipartition theorem, states that every quadratic oscillator degree of freedom will contribute with $ k_B T/2$ to the total energy,
 \begin{eqnarray}
  \label{IC_eq}\nonumber
 &&\Braket{q_i(0)} = \Braket{\dot{q}_i(0)} = 0 = \Braket{q_i(0)\dot{q}_{j}(0)}\label{ic} \\\nonumber
 &&\frac{1}{2}m_i\Braket{\dot{q}_i(0)\dot{q}_{j}(0)}= \frac{1}{2} k_BT\delta_{ij} \label{icdot}\\
 &&\frac{1}{2}m_i\omega_i^2\Braket{q_i(0)q_{j}(0)}= \frac{1}{2} k_BT\delta_{ij},\label{icpos}
  \end{eqnarray}
 where the averages are taken over the initial values with respect to the classical equilibrium density matrix of the unperturbed bath. Then, the average of Eq.~(\ref{Zeta}) vanishes, since
\begin{widetext}
\begin{equation}
\braket{   \zeta^{(0)}(t) } = - \frac{d}{dt}\sum\limits_k \frac{{C}_k}{\omega_k} \blockbrack{\cancelto{0}{\Braket{q_k^{(0)}}}\cos(\omega_kt)+\cancelto{0}{\Braket{\dot{q}_k^{(0)}}}\frac{\sin(\omega_kt)}{\omega_k}}\hspace{-0.1cm}= 0.
\end{equation}

\noindent For the two-point correlation function of the force, the non-vanishing terms are \begin{align}
\Braket{ \zeta^{(0)}(t) \zeta^{(0)}(t')}= 
\frac{d}{dt}\frac{d}{dt'}\sum\limits_k\sum\limits_{k'}\frac{{C}_k}{\omega_k}\frac{{C}_k'}{\omega_k'} &\blockbrack{\Braket{q_k^{(0)}q_{k'}^{(0)}}\cos(\omega_kt)\cos(\omega_{k'}t')+\frac{\Braket{\dot{q}_k^{(0)}\dot{q}_{k'}^{(0)}}}{\omega_k \omega_{k'}} \sin(\omega_kt)\sin(\omega_{k'}t')},
\end{align} 
 which becomes, after performing the derivatives to $t$ and $t'$,

\begin{align}
\sum\limits_k\sum\limits_{k'}{C}_k {C}_{k'}\blockbrack{\Braket{q_k^{(0)}q_{k'}^{(0)}}\sin(\omega_kt)\sin(\omega_{k'}t')+\frac{\Braket{\dot{q}_k^{(0)}\dot{q}_{k'}^{(0)}}}{\omega_k \omega_{k'}} \cos(\omega_kt)\cos(\omega_{k'}t')}.
\end{align}
\end{widetext}

Now, applying Eq.~(\ref{IC_eq}), and performing the sum over $k'$, this reduces to 
\begin{eqnarray} \label{FlucDiffGVL} \nonumber
&&\Braket{  \zeta^{(0)}(t)  \zeta^{(0)}(t') }= k_BT \sum\limits_k \frac{{C}^2_k}{m_k\omega_k^2 }\cos[\omega_k(t-t')] \\
&&= 2k_BT \int\limits_0^{\infty}\frac{d\omega}{\pi}\frac{{J}(\omega)}{\omega}\cos[\omega(t-t')], \end{eqnarray} which indeed coincides with the memory kernel in Eq.~\eqref{lvcLang}. Thus, the fluctuation-dissipation theorem holds for linear-velocity coupling.

Coming back to the equation of motion Eq.~\eqref{lvcLang}, to evaluate the frequency integral in the friction term (last term on the left-hand side), we first perform two successive partial-time integration for the friction term, 
\begin{widetext}
 \begin{eqnarray}  \label{countercancel}
  && \int\limits_0^tdt'    \int\limits_0^{\infty}\frac{d\omega}{\pi} \frac{{J}(\omega)}{\omega}\cos\blockbrack{\omega(t-t')}\dot{Q}(t') 
= \int\limits_0^tdt'  \int\limits_0^{\infty}\frac{d\omega}{\pi} \frac{{J}(\omega)}{\omega^2}\sin\blockbrack{\omega(t-t')}\ddot{Q}(t') + \int\limits_0^{\infty}\frac{d\omega}{\pi} \frac{{J}(\omega)}{\omega^2}\sin(\omega t)\dot{Q}(0)\nonumber \\ \nonumber 
&=& -  \int\limits_0^tdt'  \int\limits_0^{\infty}\frac{d\omega}{\pi} \frac{{J}(\omega)}{\omega^3}\cos\blockbrack{\omega(t-t')}\dddot{Q}(t')
+  \int\limits_0^{\infty}\frac{d\omega}{\pi} \frac{{J}(\omega)}{\omega^3} \ddot{Q}(t) - \int\limits_0^{\infty}\frac{d\omega}{\pi} \frac{{J}(\omega)}{\omega^3}\cos(\omega t)\ddot{Q}(0)
+ \int\limits_0^{\infty}\frac{d\omega}{\pi} \frac{{J}(\omega)}{\omega^2}\sin(\omega t)\dot{Q}(0).\\
\end{eqnarray}
\end{widetext}
The $\dddot{Q}$-term, i.e. the change in acceleration, is called the `jerk'. We also recognize that the second term exactly cancels the counterterm in Eq.~\eqref{lvcLang}. Then, we assume the ``superohmic bath''
\begin{equation} \label{Jnietechtohm}
{J}(\omega) =
\left\{
	\begin{array}{ll}
		\lambda \omega^3  & \mbox{if } \omega < \Omega \\
		0	  & \mbox{if } \omega > \Omega
	\end{array}
\right..
\end{equation}
The generalized Langevin equation becomes
 \begin{align}  \label{superohmicLang} \nonumber 
M\ddot{Q}(t)+V'[Q]  &-2 \lambda \int\limits_0^tdt'  \int\limits_0^{\Omega}\frac{d\omega}{\pi}  \cos\blockbrack{\omega(t-t')}\dddot{Q}(t')  \\  & =   - \bar{\zeta}^{(0)}(t),
 \end{align}  where we included both $t=0$ boundary terms, depending on $\dot{Q}(0)$ and $\ddot{Q}(0)$ into the fluctuation force, similar to what we have done before. Thus, we have 
 \begin{eqnarray} \nonumber
 &&\bar{\zeta}^{(0)}(t)  = \sum\limits_k  \frac{{C}_k}{\omega_k}  \bigg[ -\left( q_k^{(0)} - \frac{{C}_k}{m_k \omega_k^3}\dot{Q}(0) \right)   \omega_k \sin{(\omega_k t)} \\ && +\left( \dot{q}_k^{(0)} - \frac{{C}_k}{m_k\omega_k^3}\ddot{Q}(0) \right) \cos(\omega_k t) \bigg],\end{eqnarray} which does not alter the fluctuation-dissipation theorem if we take the ensemble average with respect to bath \textit{and} interaction.

Now we can evaluate the integral over $\omega$ and find the Dirac-delta distribution; hence
 \begin{align}  \label{jerkEOM}
M\ddot{Q}(t)+V'[Q] -2 \lambda  \dddot{Q}(t)   =   - \bar{\zeta}^{(0)}(t).
 \end{align}

We did  include a counterterm to the Lagrangian that, in the linear-velocity coupling case, is proportional to the square of the velocity. Since this canceled a term in Eq.~\eqref{countercancel}, this gives us the equation of motion \eqref{jerkEOM} without mass renormalization.  

\begin{widetext}
\section{Completing the square for the influence functional} \label{apD}

To perform the Gaussian integral over the bath coordinates, we need to complete the square for the bath variable $q_k$.  We can write the action as 

\begin{align} \label{act}
S^{(k)\text{E}}_{cl}=&   \frac{m_k\omega_k}{\sinh(\omega_k t)}\blockbrack{\cosh(\hbar \beta \omega_k)-1}\blockbrack{ q_k-q_k^*}^2 -\frac{m_k\omega_k}{\sinh(\hbar \beta \omega_k)} \frac{1}{\cosh{\hbar \beta \omega_k}-1} \frac{C_k^2}{2m_k \omega_k}
  \Bigl\{   
  \bigl\{
    ...   
    \bigl. \Bigl.
     \nonumber \\
&...\int\limits_0^{\hbar \beta}d\tau \int\limits_0^{\tau}d\sigma    \sinh(\omega_k \tau)\sinh\blockbrack{\omega_k(\hbar \beta - \sigma)}+\sinh(\omega_k \sigma) \sinh\blockbrack{\omega_k(\hbar \beta -\tau)}   +...
 \nonumber \\
 & ...+
  \Bigl.
  \bigl.
   \sinh(\omega_k \tau) \sinh(\omega_k \sigma)    \sinh\blockbrack{\omega_k(\hbar \beta -\tau)}        \sinh\blockbrack{\omega_k(\hbar \beta -\tau)}\bigl\}F[\dot{Q}(\tau)]F[\dot{Q}(\sigma)]
    \Bigl\}
 \nonumber \\
&-\frac{C_k^2}{m_k \omega_k \sinh(\hbar \beta \omega_k)}\int\limits_0^{\hbar \beta} d\tau \int\limits_0^{\tau} d\sigma \sinh(\omega_k \sigma)\sinh[\omega_k(\hbar \beta -\tau)]F[\dot{Q}(\tau)]F[\dot{Q}(\sigma)],
\end{align}

where
\begin{equation}
q_k^{*}=\frac{C_k}{2 m_k \omega_k} \int\limits_0^{\hbar \beta} d\tau \frac{ \sinh(\omega_k \tau)+\sinh[\omega_k (\hbar \beta-\tau)]}{\cosh(\hbar \beta \omega_k)}F[\dot{Q}(\tau)].
\end{equation}

The second term  in Eq.~\eqref{act} (comprising the two lines in the middle) is $q_n^{(0)}$, which has to be subtracted in order to complete the square; note that a factor $2$ results from adjusting the range of the $\sigma$-integral from $\{0,\hbar \beta\}$ to $\{0,\tau\}$. The last term is unchanged with respect to the Euclidean action.  

Then, we add the terms appearing under the double integrals in Eq.~\eqref{act}. To do this, we use the following trigonometric relation 

\begin{align}
 \frac{\cosh \blockbrack{a/2 - x + y}}{2\sinh(a/2)} = &\frac{1}{2 \sinh(a) \blockbrack{\cosh(a) - 1}} \Bigl\{  \blockbrack{\sinh(x) \sinh(y) + \sinh(a - x)\sinh(a - y)} + \Bigl. \nonumber \\ 
 &  \Bigl. \blockbrack{\sinh(x) \sinh(a - y) +  \sinh(y) \sinh(a - x)}  \Bigl\} +  \frac{\sinh(a - x) \sinh(y)}{\sinh(a)},
\end{align} for $a=\hbar \beta \omega_k$, $x=\omega_ks$ and $y=\omega_ku$.

The result, then, is
\begin{eqnarray} \nonumber
&&S^{(k)\text{E}}_{cl}= \frac{m_k\omega_k}{\sinh(\omega_k t)}\blockbrack{\cosh(\hbar \beta \omega_k)-1}\blockbrack{ q_k-q_k^*}^2  \\ \nonumber
&&- \frac{C_k^2}{2m_k \omega_k}
\int\limits_0^{\hbar \beta}d\tau \int\limits_0^{\tau}d\sigma   \mathcal{G}_k(\tau - \sigma) F[\dot{Q}(\tau)]F[\dot{Q}(\sigma)], \\
 \end{eqnarray} 
 
for

\begin{equation}
\mathcal{G}_k(\tau-\sigma)=\frac{\cosh\blockbrack{\omega_k\brack{\frac{\hbar \beta}{2}-(\tau-\sigma)}}}{\sinh\brack{\frac{\hbar \beta \omega_k}{2}}}.
\end{equation}
\end{widetext}

 \end{document}